\documentclass[11pt,twoside]{article}
\usepackage{asp2010}

\resetcounters

\begin{document}

\title{A sparse population of young stars in Cepheus}
\author{A. Klutsch$\,^1$, D. Montes$\,^1$, P. Guillout$\,^2$, A. Frasca$\,^3$, F.-X. Pineau$\,^2$, \\N. Grosso$\,^2$, E. Marilli$\,^3$, and J. L\'opez-Santiago$\,^1$
\affil{$^1\,$Universidad Complutense de Madrid, Departamento de Astrof\'{\i}sica, Facultad de Ciencias F\'{\i}sicas, E-28040 Madrid, Spain}
\affil{$^2\,$Observatoire Astronomique de Strasbourg, Universit\'e de Strasbourg, CNRS, UMR 7550, 11 rue de l'Universit\'e, F-67000 Strasbourg, France}
\affil{$^3\,$INAF - Osservatorio Astrofisico di Catania, via S. Sofia, 78, I-95123 Catania, Italy}}

\begin{abstract}
Once mixed in the ambient galactic plane stellar population, young stars are virtually indiscernible because neither their global photometric properties nor the presence of nearby gas can help to disentangle them from older ones. Nevertheless, the study of the \textit{RasTyc} sample revealed $4$ lithium-rich field stars displaying the same space motion, which are located within a few degrees from each other on the celestial sphere near the \textit{Cepheus-Cassiopeia} complex and at a similar distance from the Sun. Both physical and kinematical indicators show that all these stars are young, with ages in the range $10-30$\,Myr. Using multivariate analysis methods, we selected optical counterparts of ROSAT All-Sky Survey / XMM-Newton X-ray sources cross-identified with late-type stars around these $4$ young stars. Recent spectroscopic observations of this sample allowed us to discover additional lithium-rich sources. Our preliminary results showed that some of them share the same space motion as the $4$ young comoving stars. They have properties rather similar to the members of the TW~Hydrae association, although they seem to be slightly older and are located in the northern hemisphere. Nearby young stars in the field are of great importance to understand the recent local history of star formation, as well as to give new insight into the process of star formation outside standard star-forming regions and to study the evolution of circumstellar discs. 
\end{abstract}

\begin{figure}
\centering 
\includegraphics[width=12.3cm]{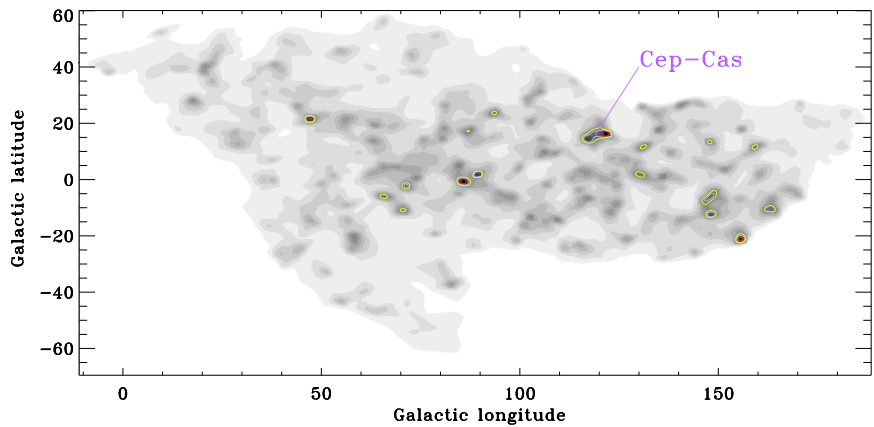}
\caption{Density map of the whole \textit{RasTyc} sources selected for our spectroscopic survey \citep{Guillout09} in galactic coordinates. We also superposed the iso-contours for density values greater than $0.32$ sources/deg$^{2}$.}
\label{Fig:RasTycDistribution}
\vspace{-0.4cm}
\end{figure}

\begin{figure}[!t]
\centering 
\includegraphics[width=12.3cm]{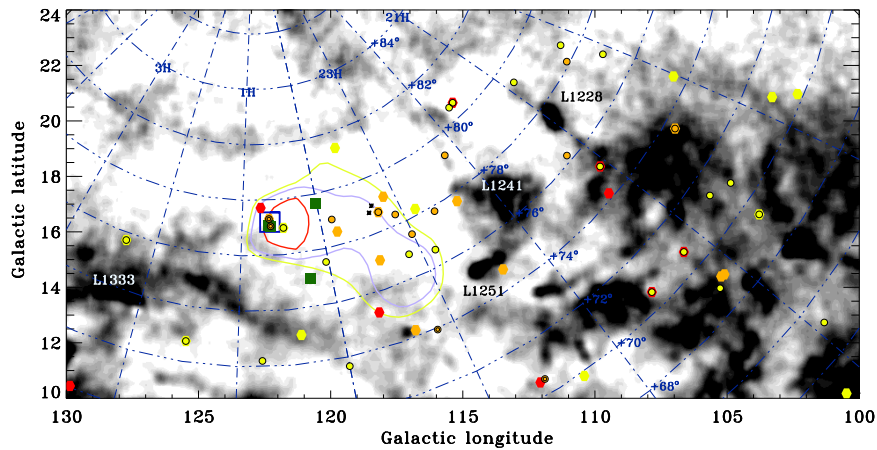}
\caption{Spatial distribution of stellar X-ray sources located near the $4$ original TTSs (filled squares) as well as the naked TTS V368 Cep and its comoving companion (asterisks), over-plotted on the \citet{Dobashi05} extinction~(A$_{v}$) map. For selected targets (hexagons) as well as \citet{Tachihara05} single stars (circles) and visual binaries (double circles), each symbol is filled in orange, red or yellow if the source displays a strong, moderate or no lithium line, respectively. The main clouds near the CO void region are labeled. We also show the density iso-contours for the whole \textit{RasTyc} sources as in Fig.~\ref{Fig:RasTycDistribution} and the locus of an unusual concentration of at least $7$ young stars (big open square) outside of SFRs.}
\label{Fig:CoMVMap}
\vspace{-0.4cm}
\end{figure}

\section{Introduction on stellar X-ray sources in the solar neighborhood}

Most stars detected by the ROSAT mission are younger than $1$\,Gyr \citep[e.g.][]{Motch97}. Taking into account this property, \citet{Guillout99} cross-correlated the ROSAT All-Sky Survey (RASS) with the Tycho catalogue creating the largest ($\approx 14 000$ active stars) and most comprehensive set of late-type stellar X-ray sources, the so-called \textit{RasTyc} sample. This stellar population can be used as a tracer of local young structures \citep{Guillout98}. Presently, nine nearby ($30-150$\,pc) young ($5-70$\,Myr) associations are already identified in the southern hemisphere (see the reviews by \citealt{ZS04}, and \citealt{Torres08}), e.g. the TW Hydrae association (TWA) around TW Hya \citep{Gregorio-Hetem92, Kastner97}. In particular, the SACY survey \citep{Torres06,Torres08} has discovered several members of such association. This sample represents a sub-sample of the \textit{RasTyc} population in the southern hemisphere. Our spectroscopic observations of the optically bright \textit{RasTyc} sources, seen in the northern hemisphere, allowed us to identify 5 young stars \citep{Guillout09}, but none is near the largest over-density of our sources (Fig.~\ref{Fig:RasTycDistribution}). However, the sky density is more uniform and about one order of magnitude lower, on average, than in the SACY survey. This is consistent with the significant asymmetry in the all-sky \textit{RasTyc} distribution with respect to the galactic plane, already shown by \citet{Guillout98}. 

\section{Discovery of four comoving T Tauri stars}

From the analysis of our first spectroscopic observations of optically faint \textit{RasTyc} stars, we discovered an unusual group \citep{Klutsch08} of $4$ lithium-rich stars (filled squares on Fig.~\ref{Fig:CoMVMap}), towards the \textit{Cepheus-Cassiopeia} (Cep-Cas) complex.~Although this sky area is rich in CO molecular regions \citep{Dame01} and dark clouds \citep{Dobashi05}, these stars are projected several degrees off-clouds in front of a region devoid of interstellar matter, which corresponds precisely to the sky area with the highest density of \textit{RasTyc} sources (Figs.~\ref{Fig:RasTycDistribution} and \ref{Fig:CoMVMap}). 

\begin{figure}[!t]
\hspace{-0.975cm}
\centering 
\includegraphics[width=4.9cm]{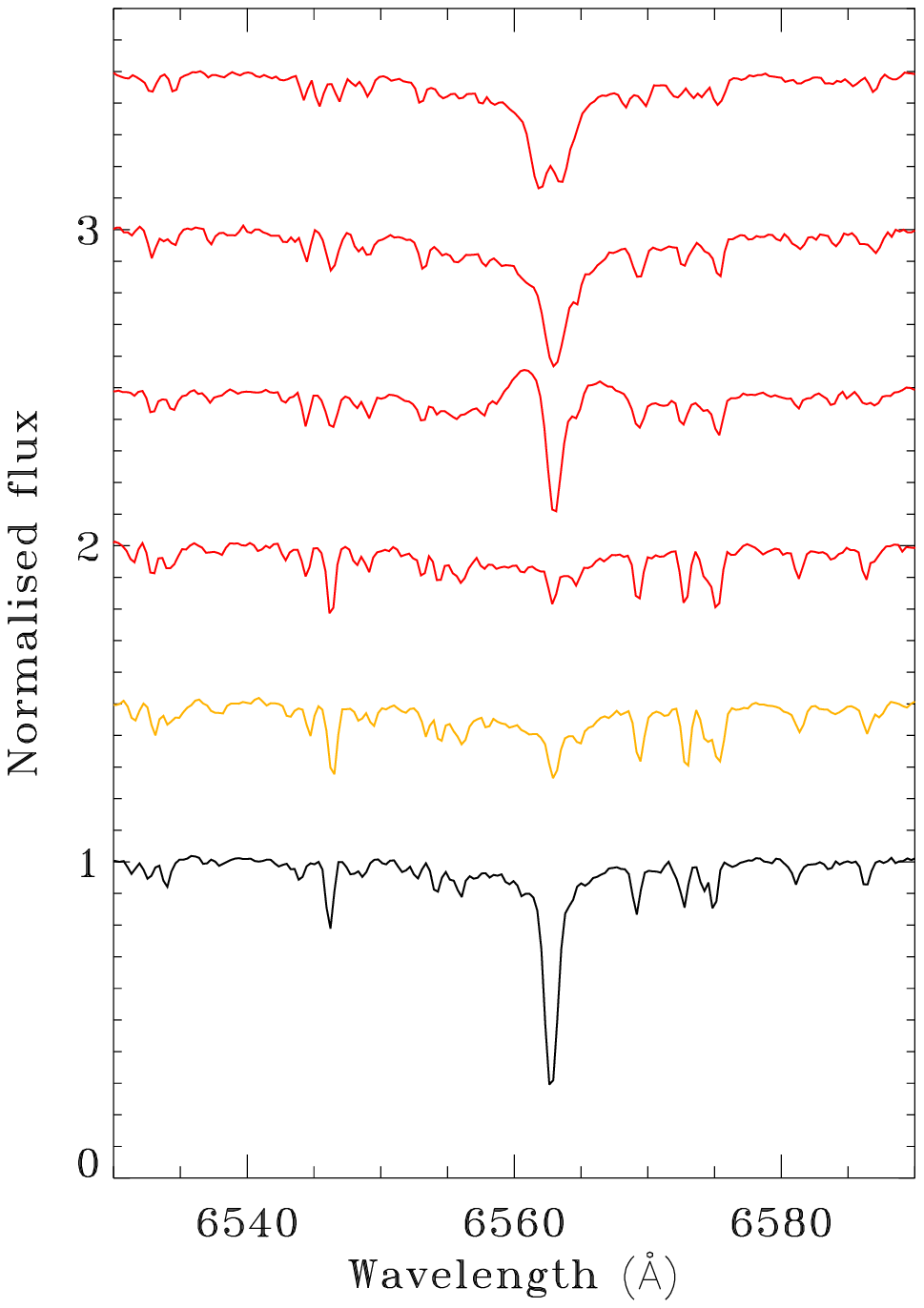}
\hspace{-1.55cm}
\includegraphics[width=4.9cm]{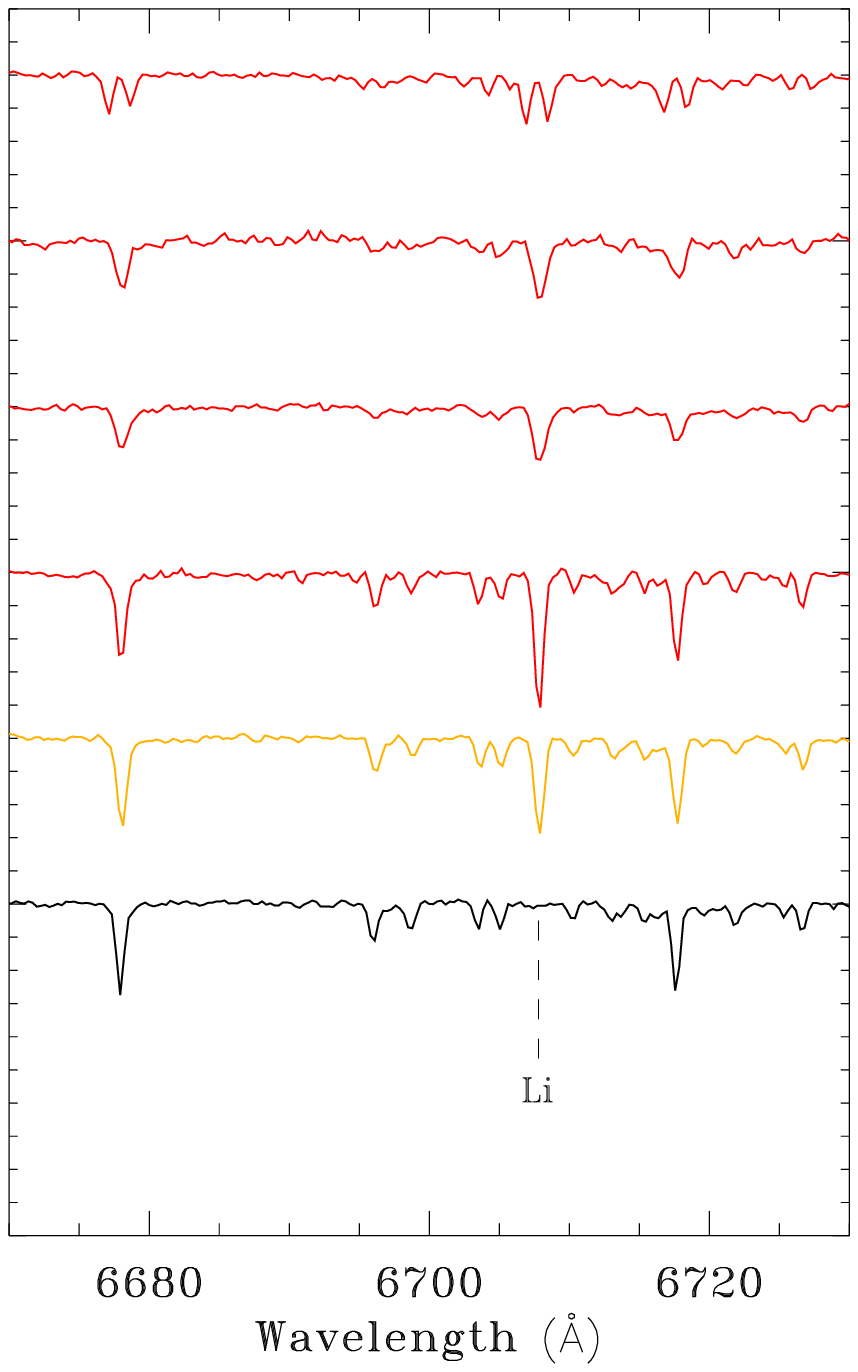}
\hspace{-0.5cm}
\includegraphics[width=6.1cm]{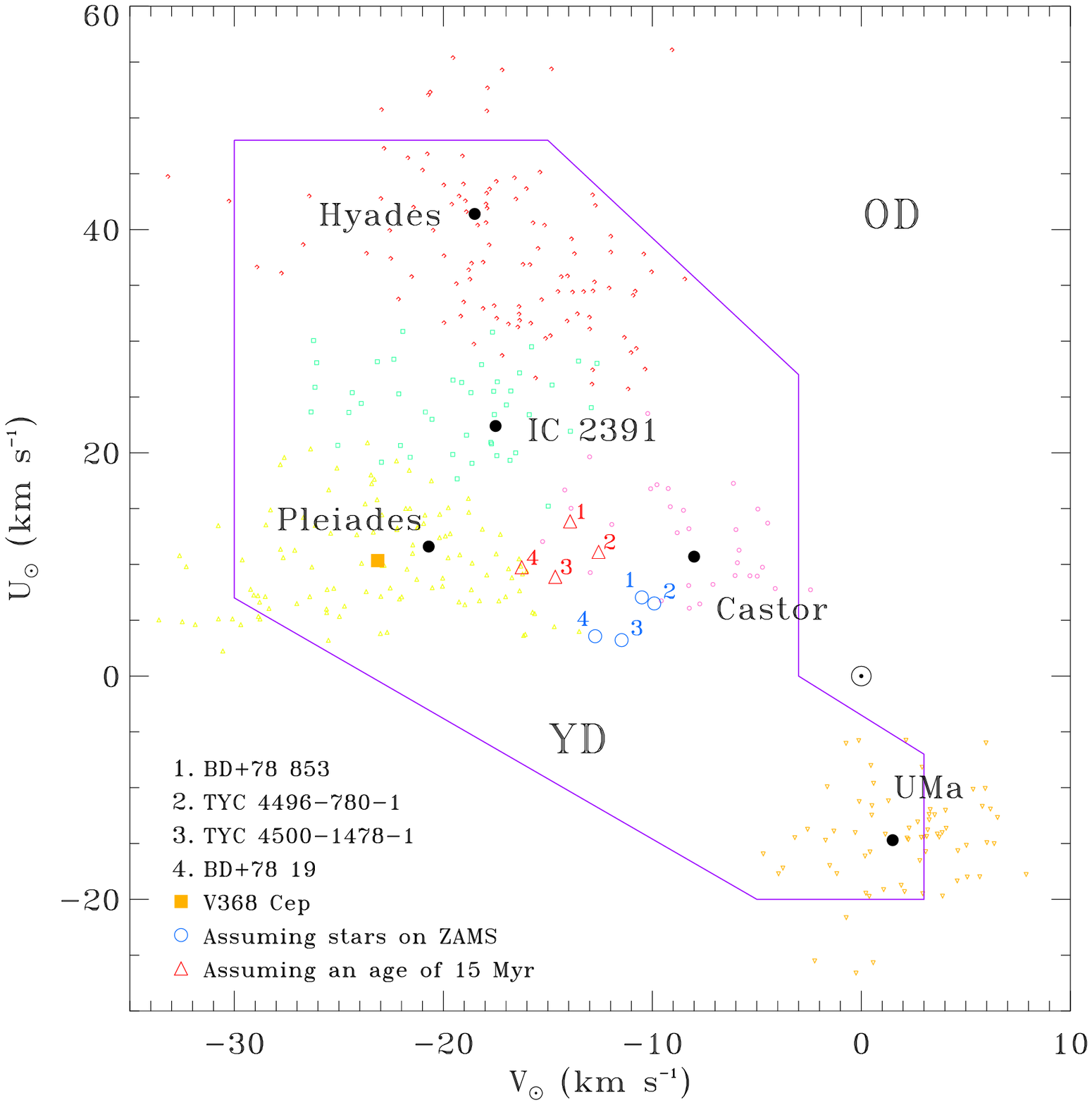}
\caption{\textit{Left}: IDS spectra (shifted vertically) centered in $60~\AA$-wide region around the H$\alpha$ (left box) and lithium (right box) lines for the 4 comoving TTSs (red) and the naked TTS V368 Cep (orange). We also displayed the spectrum of a ``non-active'' standard star (black). \textit{Right}: $U-V$ kinematic diagram of our 4 comoving TTSs. The average velocity components (dots) of some young stellar kinematic groups (SKG) and those of some of their late-type members are also plotted. The loci of the young-disc (YD) and old-disc (OD) populations are also marked.}
\label{Fig:OldYoungStars_Plot}
\end{figure}

\begin{figure}[!t]
\centering
\includegraphics[height=4.34cm]{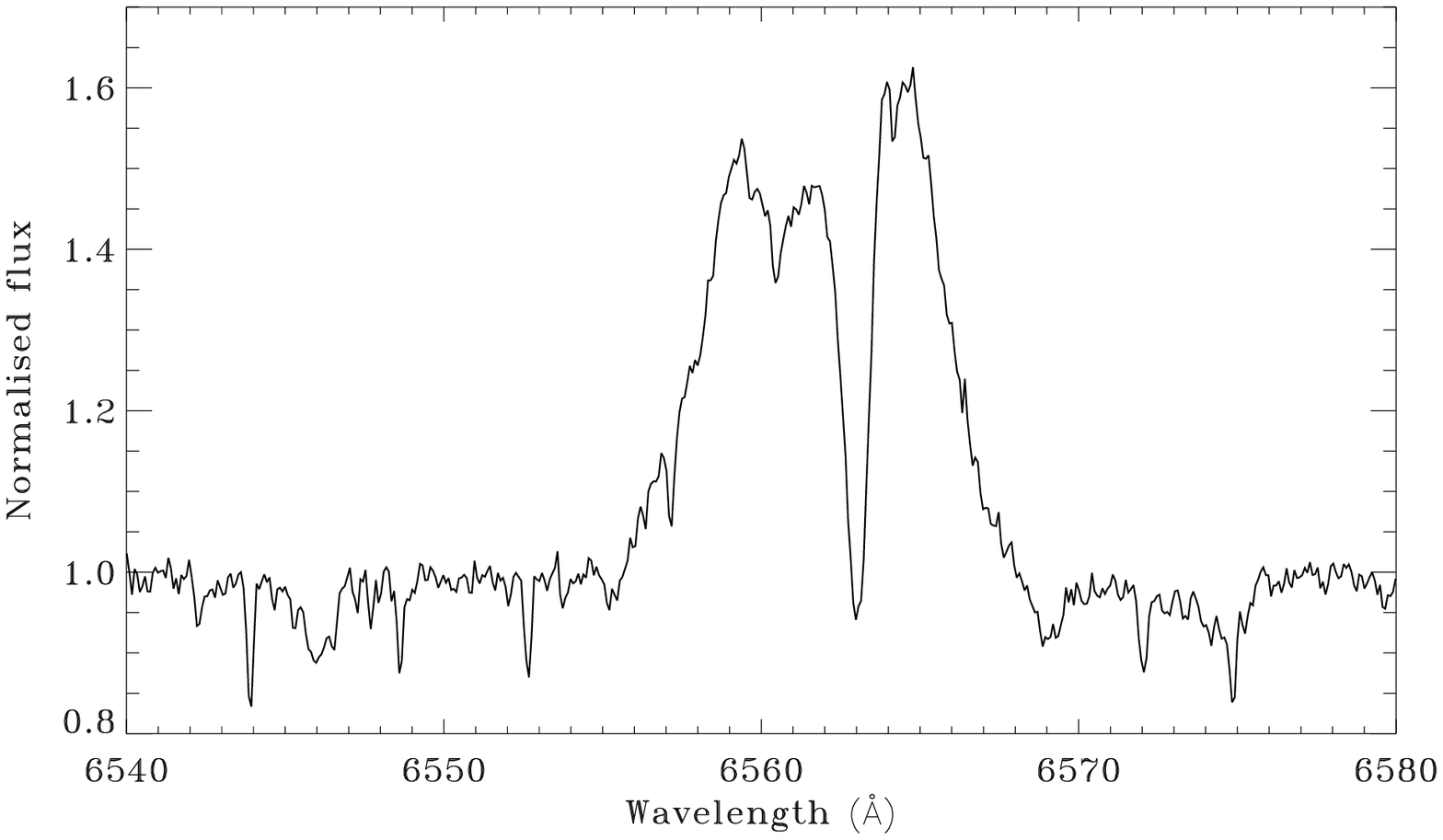}
\hspace{-0.4cm}
\includegraphics[height=4.45cm]{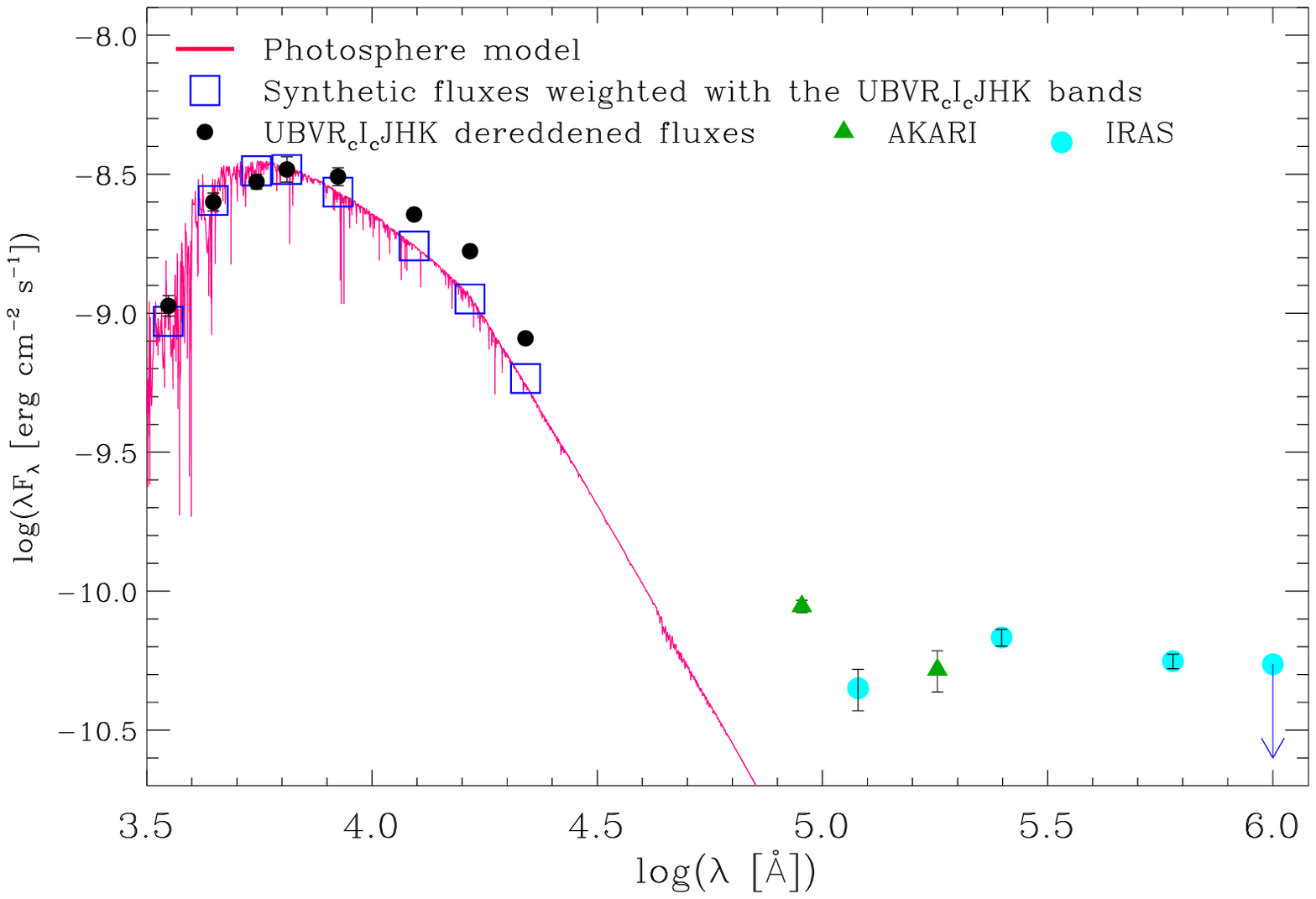}
\caption{\textit{Left}: FOCES spectrum of TYC\,4496-780-1 displaying a strong H$\alpha$ emission. \textit{Right}: Its spectral energy distribution showing a near- and far-infrared excess, a typical signature of an accretion disc (\citealt{Guillout10a}).}
\label{Fig:TYC4496}
\end{figure}

\smallskip
They have all typical spectral signatures of young stars \citep{Guillout10a} as \textit{i)} a H$\alpha$ emission or a filled-in profile (left panel of Figs.~\ref{Fig:OldYoungStars_Plot} and \ref{Fig:TYC4496}), \textit{ii)} a strong lithium~absorption line (Fig.~\ref{Fig:OldYoungStars_Plot}, middle panel) corresponding to a lithium abundance close to the primordial one, and \textit{iii)} a X-ray luminosity (L$_{X}$) of $\sim 10^{30.4}$\,erg s$^{-1}$ (within $0.2$ dex) which is similar to that observed for weak-line T Tauri stars (WTTS) in Taurus-Auriga-Perseus star-forming regions (SFRs). Only TYC\,4496-780-1, the star with a strong H$\alpha$ emission (Fig.~\ref{Fig:TYC4496}, left panel), displays a near- and far-infrared excess (Fig.~\ref{Fig:TYC4496}, right panel). This feature is typical of class II infrared sources, i.e. T Tauri stars (TTSs) still surrounded by an accretion disc. Because of the lack of relevant infrared excess, the $3$ other sources are likely WTTS or post-T Tauri stars whose gaseous discs have already been dissipated as it is observed in most stars with an age between $10-70$\,Myr. 

\smallskip
Unfortunately, their Tycho parallaxes are useless and one must rely on photometric distance estimations. To cover a wide range of possibilities, we have estimated for each star two distances. We considered that the \textsl{lower limit} is $80-100$\,pc if stars are on the zero-age main sequence (ZAMS) and that the \textsl{upper limit} is $130-180$\,pc assuming a stellar age of $15$ Myr. Using a lower age would be in contradiction to our photometric observations showing that no stars suffer major interstellar extinction \citep{Guillout10a}. Fig.~\ref{Fig:OldYoungStars_Plot} (right panel) shows that, whatever the distance is, our $4$ stars seem to share the same kinematics (within a few km s$^{-1}$) proving that they form a homogeneous group with a common origin and they are unrelated with the naked TTS V368 Cep.

\smallskip
We also consider a possible link with the Cep-Cas complex. We can not exclude that these stars are runaway objects originated in L1251, L1241 or L1228, but their high-escape velocity and their similar space motion cast doubt on this hypothesis. The more plausible explanation for the formation of these TTSs is the in-situ model. 

\section{Search for new young comoving candidates}

\begin{figure}[!ht]
\hspace{-0.5cm}
\centering 
\includegraphics[height=5.6cm]{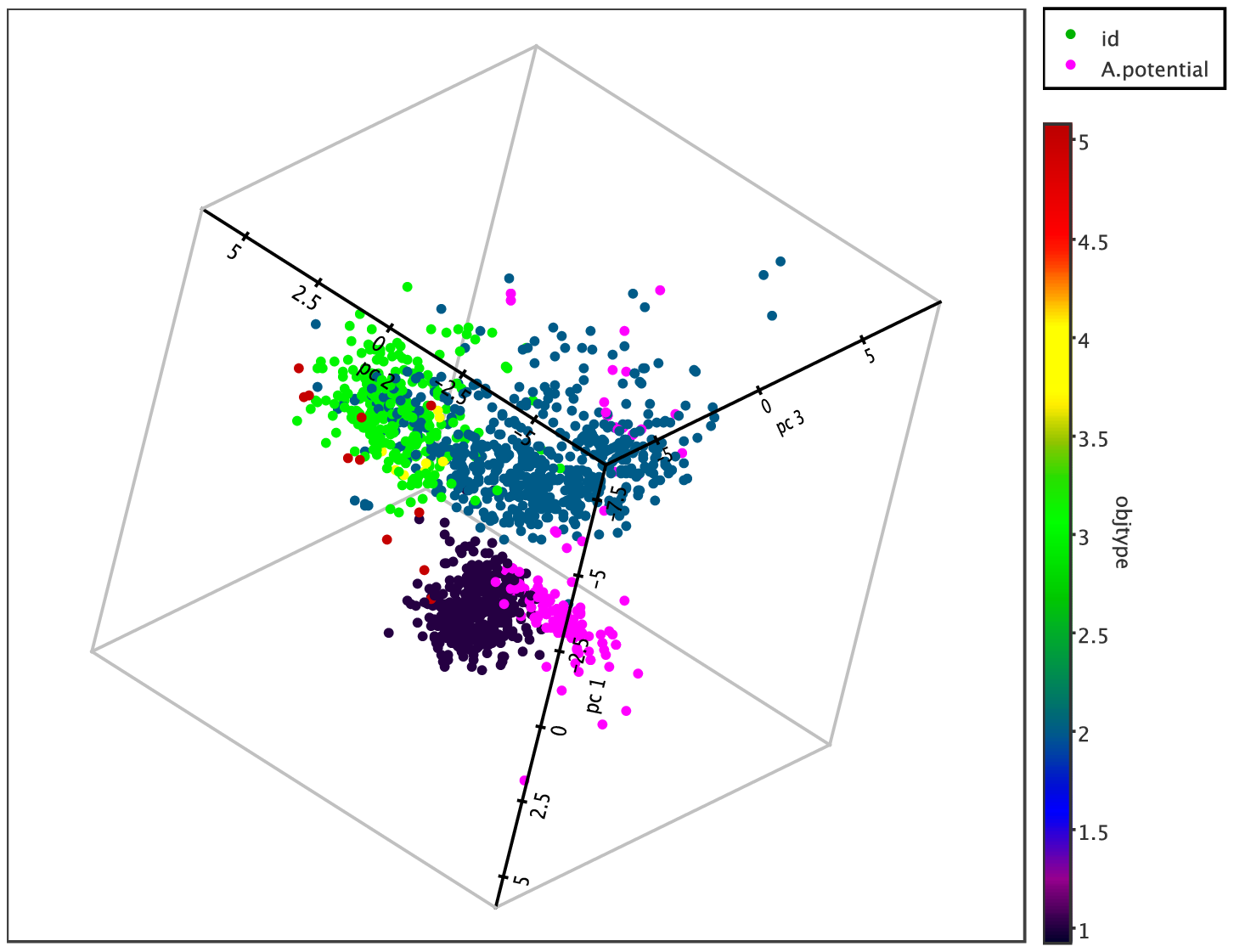}
\hspace{-0.9cm}
\includegraphics[height=5.6cm, bb= 0 0 365 374,clip=]{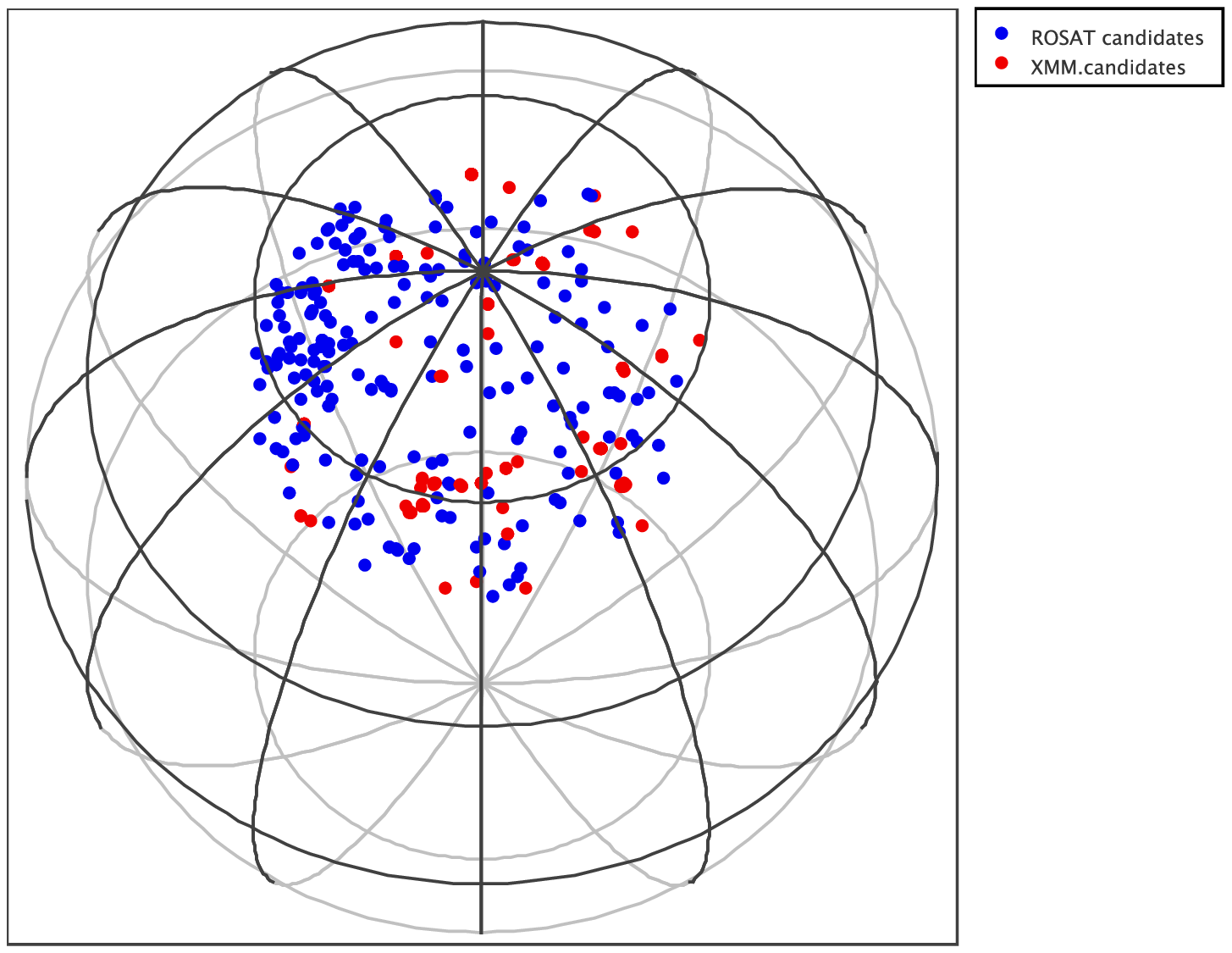}
\caption{\textit{Left}: Result of the multivariate analysis applied on XMM X-ray sources that are formerly identified with quasars (green symbols), galaxies (blue symbols) and stars (other symbols). These methods maximize the separation of the various classes of X-ray emitters allowing to select a priori, with a high probability, young star candidates (pink symbols). \textit{Right}: Sky position of XMM / RASS young star candidates (red and blue circles, respectively) meeting all our selection criteria.}
\label{Fig:YoungStars_ACP}
\end{figure}

Selecting an appropriate sample of targets is the major difficulty for searching other comoving young stars in this sky area. We picked out counterparts of XMM-Newton (XMM) X-ray sources cross-identified with the 2MASS catalog using a likelihood ratio approach and an original way to estimate the rate of spurious association (\citealt{Pineau09}; Pineau et al. in prep.). We discarded all associations with a separation greater than $5$\,arcsec and a probability of identification lower than $0.7$, and correlated the remaining sources with the GSC2 catalog. We kept all GSC2 sources lying at a distance not exceeding $5$\,arcsec from an XMM source and $1$\,arcsec from a 2MASS source. The distance limit of RASS X-ray sources with both GSC2 and 2MASS catalogs is $15$\,arcsec. We then performed a principal component analysis (PCA), taking into account measurements errors \citep{Pineau09}, on $19$ parameters built up by linear combinations of flux ratios (from the X-ray to the near-IR domain), color
indices, and X-ray hardness ratios. We used a mean shift made on the $3$ first principal components of the PCA for disentangling the stellar population from the extragalactic component (galaxies and quasars) also emitting in X-ray (Fig.~\ref{Fig:YoungStars_ACP}, left panel). All the stars with the same maximum local as the young stars known in Simbad are classified as the \textit{young stars candidates}. 

\smallskip
We observed all the candidates which are in a region $30\deg$ wide encompassing our $4$ comoving TTSs and meet the following selection criteria: \textit{i)} late-type stars -- $B-V > 0.6$, \textit{ii)} faint -- $V > 10$ mag, \textit{iii)} X-ray luminous -- L$_{X} > 10^{29}$ erg s$^{-1}$, and \textit{iv)} within $170$ pc of the Sun (Fig.~\ref{Fig:YoungStars_ACP}, right panel). To characterize these stars, we conducted intermediate- and high-resolution spectroscopic observations on $2-3$\,m class telescopes during the semester $2009$B with INT/IDS at the \textit{Roque de Los Muchachos Observatory} at La Palma (Spain), 2.2m/FOCES at \textit{the Centro Astron\'omico Hispano Alem\'an} (CAHA) at Calar Alto (Spain), and T193/Sophie at the \textit{Observatoire de Haute Provence} (France). Further observations are scheduled in November $2010$. 

\begin{table}[!t]
\caption{Preliminary classification of observed stars for each instrument}
\smallskip
\begin{center}
{\footnotesize
\begin{tabular}{lcccccc}
\tableline
\noalign{\smallskip}
 & & \multicolumn{4}{c}{Number (and fraction) of stars with:} &  Spectroscopic~~ \\ 
\noalign{\smallskip}
\raisebox{1.5ex}[0cm][0cm] {Spectrograph}  & \raisebox{1.5ex}[0cm][0cm] {N} & Li strong & Li moderate & H$\alpha$ emission & H$\alpha$ filled & binaries~~ \\ 
\noalign{\smallskip}
\tableline
\noalign{\smallskip}
IDS       & $76$ & ~$8$~ ($11\,\%$)  & ~$9$~ ($12\,\%$)  & $42$ ($55\,\%$)  & $13$ ($17\,\%$)  & $19-21$~($25-28\,\%$) \\ 
FOCES & $31$ & $11$ ($35\,\%$)  & ~$9$~ ($29\,\%$)  & ~$5$~ ($16\,\%$)  & ~$7$~ ($23\,\%$)  & ~$8~-10$~($26-32\,\%$) \\ 
Sophie  & $18$ & ~$6$~ ($33\,\%$)  & ~$5$~ ($28\,\%$)  & ~$4$~ ($22\,\%$)  &$11$ ($61\,\%$)  &  ~$5~-~6$~~($28-33\,\%$) \\
\noalign{\smallskip}
\tableline 
\noalign{\smallskip}
Total & $125$  & $25$ ($20\,\%$)  & $23$ ($18\,\%$)  & $51$ ($41\,\%$)  & $31$ ($25\,\%$)  & $32-36$ ($26-30\,\%$) \\
\noalign{\smallskip}
\tableline 
\end{tabular}
\label{Tab:Results}
}
\end{center}
\vspace{-0.4cm}
\end{table}

\begin{table}[!t]
\caption{List of young star candidates displaying a strong lithium line}
\smallskip
\begin{center}
{\footnotesize
\begin{tabular}{lccccc}
\tableline
\noalign{\smallskip}
 & $\alpha$ (2000) & $\delta$ (2000) & V &  \\ 
\raisebox{1.5ex}[0cm][0cm] {Name} & (h m s) & ($\deg ~\arcmin ~\arcsec$) & (mag) & \raisebox{1.5ex}[0cm][0cm] {Ref.} \\
\noalign{\smallskip}
\tableline
\noalign{\smallskip}
1RXS\,J003904.2+791912$\,^{\dag,\star,\ddag}$ &	00 39 03.55 &	+79 19 19.2 &	13.4 & 1,2\\ 
1RXS\,J020434.1+675635 &	02 04 32.97 &	+67 56 31.1 &	12.3 & 2\\ 
1RXS\,J030016.2+722537 &	03 00 14.78 &	+72 25 40.9 &	10.8 & 2\\ 
1RXS\,J044912.7+773719 &	04 49 15.32 &	+77 37 12.3 &	11.0 & 2\\ 
1RXS\,J071537.9+764648 &	07 15 32.36 &	+76 47 08.4 &	10.7 & 2\\ 
1RXS\,J071743.1+764416 &	07 17 42.40 &	+76 44 20.4 &	11.3 & 2\\ 
1RXS\,J181048.9+701601 &	18 10 49.98 &	+70 16 10.0 &	10.7 & 2\\ 
1RXS\,J185131.1+584258 &	18 51 30.99 &	+58 42 58.0 &	11.1 & 2\\ 
1RXS\,J191519.4+660830 &	19 15 16.19 &	+66 08 51.6 &	10.7 & 2\\ 
1RXS\,J194206.0+654655 &	19 42 05.95 &	+65 46 49.9 &	10.9 & 2\\ 
1RXS\,J200245.4+592015 &	20 02 44.63 &	+59 20 16.4 &	10.8 & 2\\ 
1RXS\,J202513.8+733638 &	20 25 15.40 &	+73 36 33.3 &	10.7 & 1,2\\ 
1RXS\,J210444.0+522326 &	21 04 44.33 &	+52 23 25.6 &	12.1 & 2\\ 
1RXS\,J210751.6+690923 &	21 07 51.95 &	+69 09 23.4 &	13.3 & 2\\ 
1RXS\,J211000.9+603644 &	21 10 01.98 &	+60 36 45.4 &	11.9 & 2\\ 
1RXS\,J211114.9+614436 &	21 11 16.69 &	+61 44 38.8 &	11.8 & 2\\ 
1RXS\,J211627.5+575501 &	21 16 25.59 &	+57 54 57.1 &	11.3 & 2\\ 
1RXS\,J212606.8+555026 &	21 26 06.12 &	+55 50 24.5 &	11.5 & 2\\ 
1RXS\,J222257.8+744322 &	22 22 58.98 &	+74 43 21.0 &	12.8 & 2\\ 
1RXS\,J225641.1+593031 &	22 56 40.95 &	+59 30 37.0 &	11.5 & 2\\ 
1RXS\,J225855.7+673117 &	22 58 53.92 &	+67 31 18.5 &	11.9 & 2\\ 
1RXS\,J230822.7+790829$\,^{\dag}$ &	23 08 22.21 &	+79 08 23.7 &	12.8 & 2\\ 
1RXS\,J231616.5+784156$\,^{\dag}$ &	23 16 18.34 &	+78 41 56.0 &	11.8 & 1,2\\ 
1RXS\,J232647.5+770304$\,^{\dag}$ &	23 26 49.43 &	+77 03 06.1 &	13.2 & 2\\ 
1RXS\,J234920.7+783000$\,^{\dag}$ &	23 49 20.06 &	+78 29 46.8 &	11.6 & 2\\ 
\tableline 
\noalign{\smallskip}
\end{tabular}
\begin{tabular}{l}
Notes: $^{\dag}$ towards the CO Cepheus void; $^{\star}$ in the smaller stellar concentration; $\,^{\ddag}$ visual binary\\
References: $^{1}$ \citet{Tachihara05}; $^{2}$ this study\\
\end{tabular}
\label{Tab:LiRich_Star}
}
\end{center}
\vspace{-0.5cm}
\end{table}

\begin{figure}[h]
\hspace{-0.975cm}
\centering 
\includegraphics[width=4.25cm]{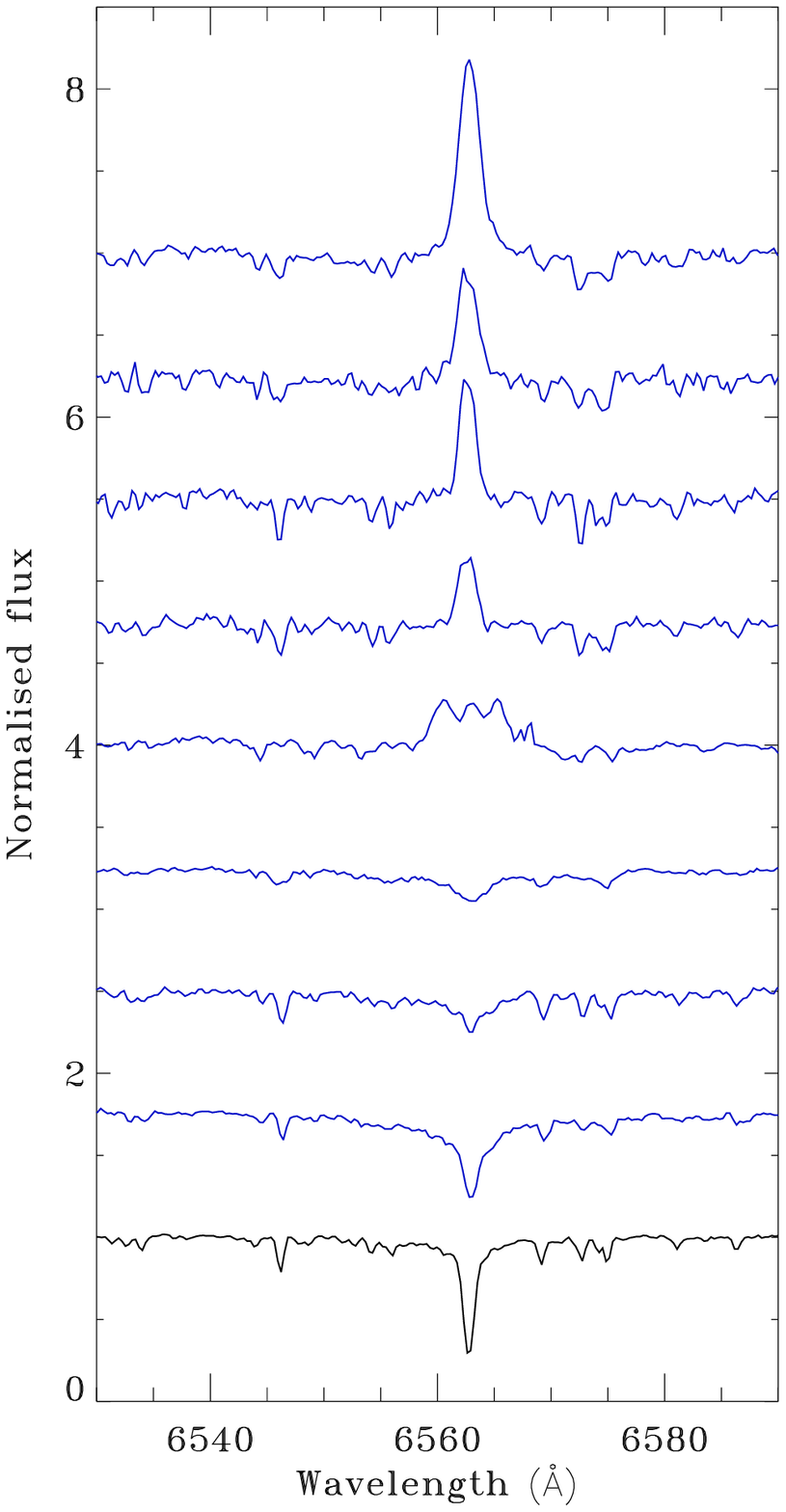}
\hspace{-1.365cm}
\includegraphics[width=4.25cm]{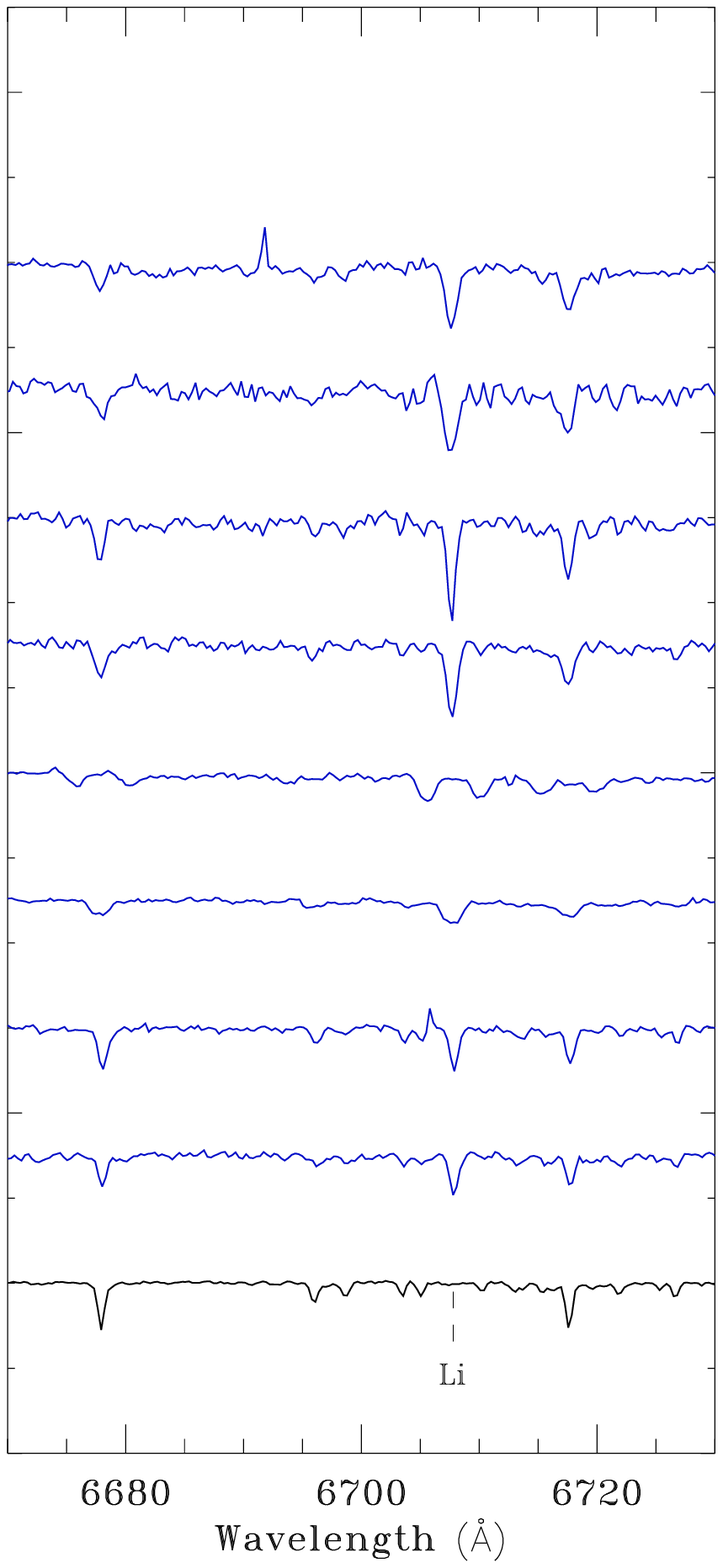}
\hspace{-0.5cm}
\includegraphics[width=7.175cm]{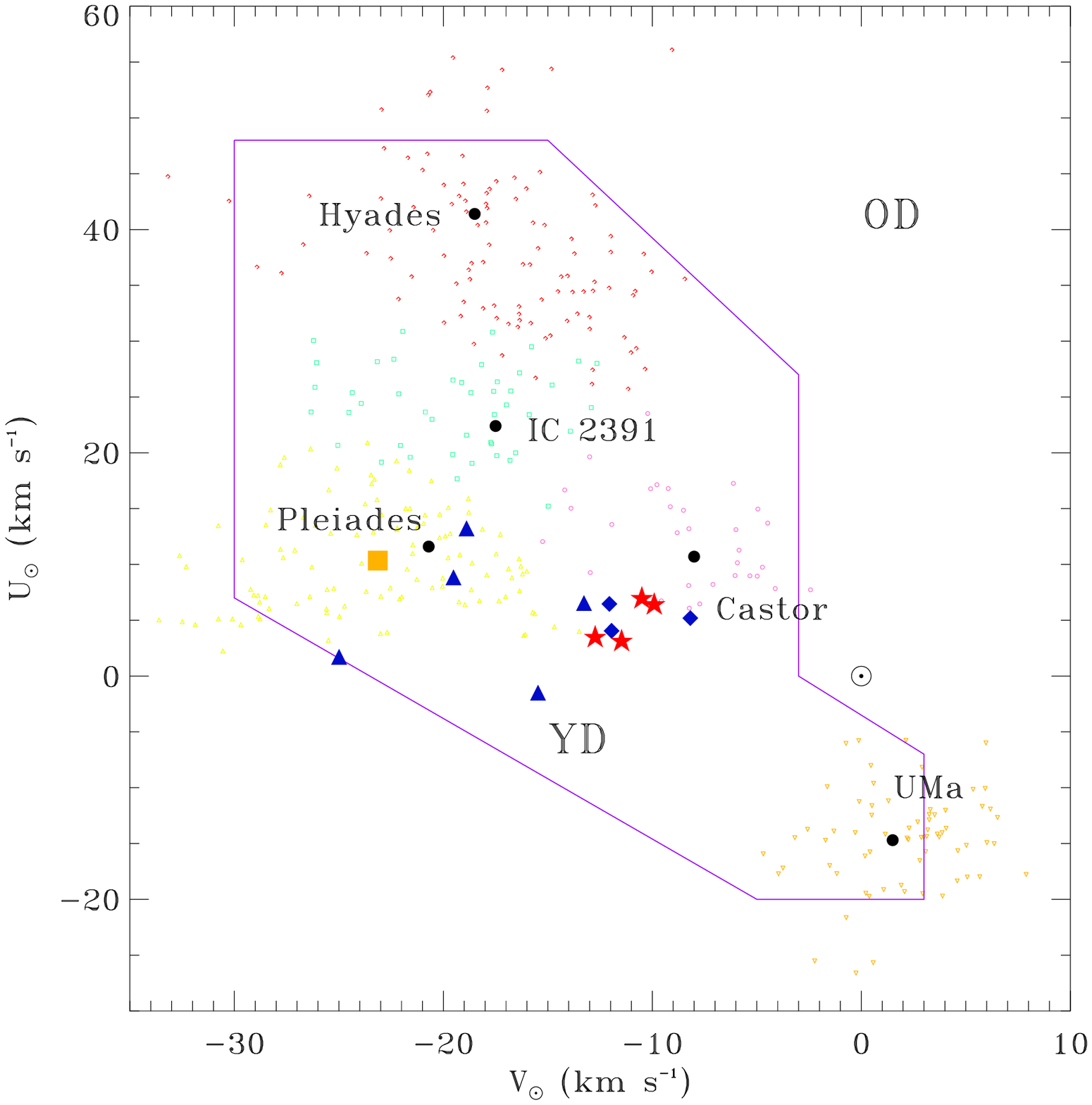}
\caption{As for Fig~\ref{Fig:OldYoungStars_Plot} but for $8$ of our candidates displaying a strong lithium line. On the right panel, the V and U velocity components are displayed with triangles and diamonds for the single stars and the spectroscopic systems, respectively. The positions of the $4$ original TTSs are also plotted as star symbols.}
\label{Fig:YoungStars_Plot}
\vspace{-0.4cm}
\end{figure}

\begin{figure}[!t]
\centering 
\includegraphics[width=7.0cm]{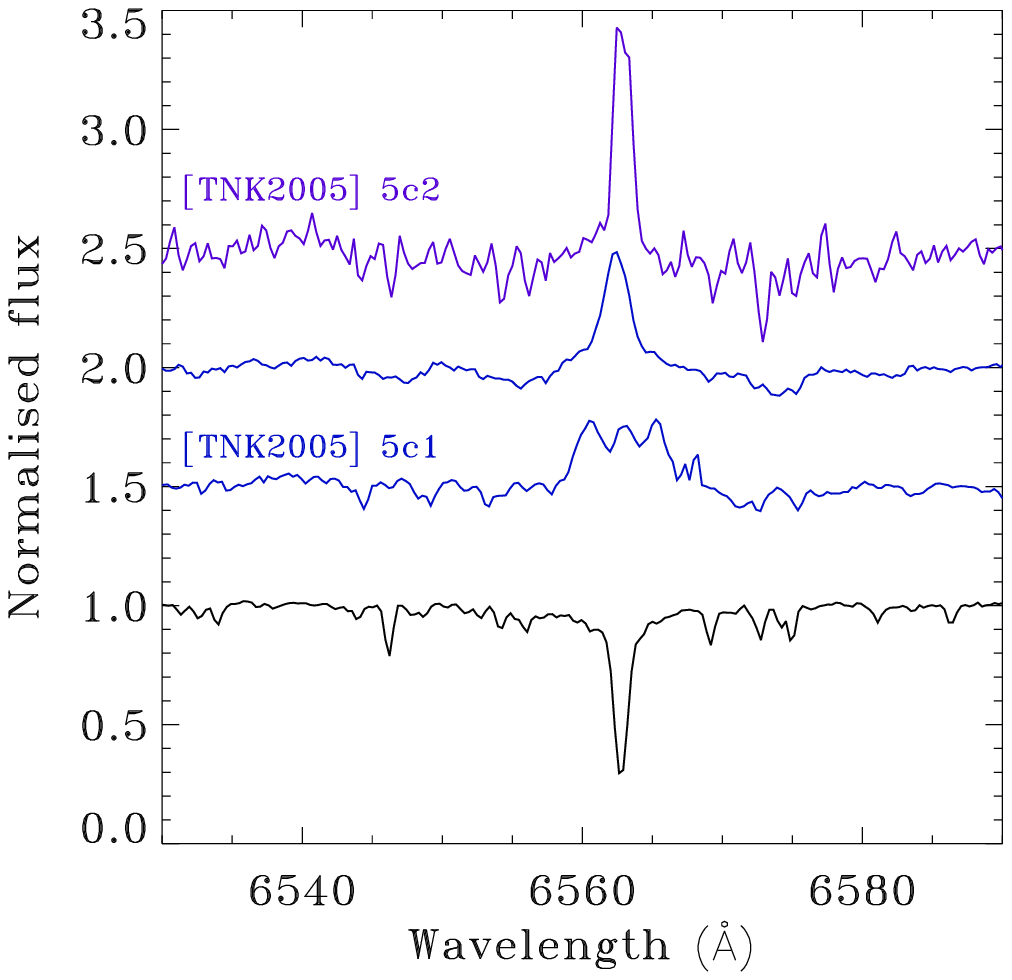}
\hspace{-2.1cm}
\includegraphics[width=7.0cm]{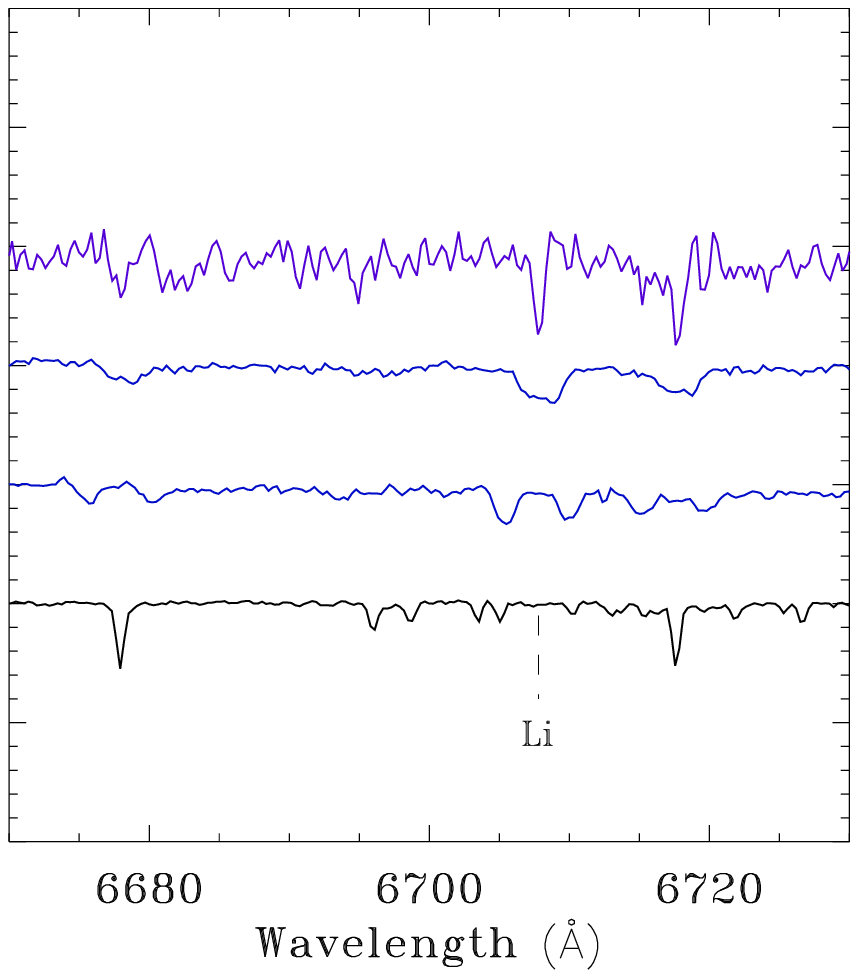}
\caption{H$\alpha$ and lithium spectra of the two visual components of [TNK2005] $5$. We plot two spectra of the brightest component taken in different configurations. We also acquired a low signal-to-noise spectrum of the faintest component. }
\label{Fig:triple_System}
\vspace{-0.4cm}
\end{figure}

\smallskip
We started a classification of these candidates according to their spectral features (Table~\ref{Tab:Results}). In particular, we identified $25$ sources displaying a strong lithium line similar to that of our original TTSs (Table~\ref{Tab:LiRich_Star}). Out of $8$ lithium-rich stars discovered with IDS (Fig.~\ref{Fig:YoungStars_Plot}), $4$ are good young comoving candidates, while $3$ turn out to be spectroscopic binaries. Spectra of the remaining $17$ are in the reduction/analysis phase. Moreover three of them and one of the original TTSs were already identified as WTTS (orange filled circles on Fig.~\ref{Fig:CoMVMap}) by \citet{Tachihara05}. All the stars studied in the latter paper will also be included in our future observing runs for determining their radial velocity (not provided by \citeauthor{Tachihara05}) and kinematics in order to establish a possible link with our $4$ comoving TTSs. 

\smallskip
We also noticed that more than $65\,\%$ of candidates display an emission or filled-in H$\alpha$ profile (Table~\ref{Tab:LiRich_Star}). With about $26\,\%$ of double-lined spectroscopic binaries in our sample, we found a larger fraction (two orders of magnitude) than that found by \citet{Guillout09} in the optically bright \textit{RasTyc} sample. 

\section{New young stars in the CO Cepheus void}

On Fig.~\ref{Fig:CoMVMap}, we showed the spatial distribution of the selected stars (hexagon symbols) around the $4$ comoving TTs. Presently, $15$ stellar X-ray sources located in the CO Cepheus void, 8 of which were already discovered by \citeauthor{Tachihara05}, are rich in lithium. Out of them, $3$ (including $2$ spectroscopic systems) are young comoving candidates. The discovery of such a number of T Tauri stars and new stars displaying a strong lithium line (i.e. T Tauri candidates) in this region is more easily explicable by the in-situ model than by the runaway hypothesis.

\smallskip
We also detected an unusual concentration (big open square on Fig.~\ref{Fig:CoMVMap}) of at least $7$ lithium-rich stars formed by two of the original TTSs (the spectroscopic binary, RasTyc0039+7905 and the visual binary, RasTyc0038+7903 or [TNK2005]\footnote{Acronym of the \citet{Tachihara05} sources created by Simbad, the CDS database.} $4$) and a binary-pair candidate, [TNK2005] $5$. During one night, both components of this last source were positioned simultaneously on the slit. We observed the brightest component as a double-lined binary and acquired a spectrum of the faintest star (Fig.~\ref{Fig:triple_System}). \citeauthor{Tachihara05} were already detected this faint star from their photometric observations and classified it as a M2-type companion candidate. This star displays a prominent H$\alpha$ emission and a dip lithium absorption. Moreover, its radial velocity is similar to the barycentric velocity of the binary system. Thus, [TNK2005] $5$ turns to be a triple system composed of a close inner binary (the brightest component) plus a tertiary component in a long-period orbit. These sources are close to a faint, low density cloud \citep{Tachihara05} which could be the remnant of the parent cloud already dissipated. 

\smallskip
These preliminary results suggest a link between some young stars of this region. If this will be confirmed, the mass of the molecular cloud will exceed $800$ M$_{\odot}$ expected by \citet{Tachihara05} and the most plausible of their hypothesis will be that these stars were affected by an yet unknown supernova shock.

\section{Conclusions and perspectives}

We have presented preliminary results of a spectroscopic survey of optical counterparts of X-ray sources in the Cepheus region. Several lithium-rich stars (likely WTTS) have discovered in a region devoid of dense molecular clouds, the so-called Cepheus void. Among them, TYC\,4496-780-1 seems to be a class II young infrared source, i.e. a TTS still surrounded by an accretion disc. Presently, few of these sources have been found outside of SFR's cores (e.g. TW Hya). All the stars discussed have properties similar to the TW Hya association, although they seems to be slightly older and are located in the northern hemisphere. Some of them also form an unusual concentration of at least $7$ lithium-rich stars in a small sky area. The runaway hypothesis is highly improbable for explaining the formation of this homogeneous comoving group because of their kinematical properties and the identification of a large number of new T Tauri candidates in this region. This raises the question of the in-situ star-formation scenario in low-mass cloud environments (as in many other SFRs). Afterwards the Gaia mission will certainly shed light on this issue and on the origin of this group which could be related to the Cep-Cas complex. 

\smallskip
Further details of this work will be presented in a future publication.

\acknowledgements 
This work is supported by Universidad Complutense de Madrid, the Spanish \textit{Ministerio de Ciencia e Innovaci\'on} (MICINN) under grants AYA2008-00695 and AYA2008-06423-C03-03, and AstroMadrid (CAM S2009/ESP-1496). 
Part of this study is supported by the Italian \textit{Ministero dell'Istruzione, Universit\`a e  Ricerca} (MIUR), and the \textit{R\'egion Alsace}. 

\bibliography{Klutsch_A}
\end{document}